%% file: DT_CB_Learning.tex
\documentclass[conference]{IEEEtran}
\usepackage[top=0.7in, bottom=0.96in, left=0.65in, right=0.65in]{geometry} 
\input{input.tex}

\usepackage{epstopdf}
\usepackage{enumerate}
\usepackage{float}
\usepackage{color}
\usepackage{makeidx}
\usepackage{bm}
\usepackage{cleveref}
\usepackage{url}
\usepackage{subfigure}
\usepackage{balance}

\DeclareMathOperator*{\argmin}{arg\,min}
\DeclareMathOperator*{\argmax}{arg\,max}

\begin{document}
    \bstctlcite{IEEEexample:BSTcontrol}

    \title{Digital Twin Aided Millimeter Wave MIMO: \\ Site-Specific Beam Codebook Learning}

    \author{Hao~Luo~and~Ahmed~Alkhateeb\\School of Electrical, Computer, and Energy Engineering, Arizona State University\\Email: \{h.luo, alkhateeb\}@asu.edu}

    \maketitle

    \begin{abstract}	
        Learning site-specific beams that adapt to the deployment environment, interference sources, and hardware imperfections can lead to noticeable performance gains in coverage, data rate, and power saving, among other interesting advantages. 
        This learning process, however, typically requires a large number of active interactions/iterations, which limits its practical feasibility and leads to excessive overhead. 
        To address these challenges, we propose a digital twin aided codebook learning framework, where a site-specific digital twin is leveraged to generate synthetic channel data for codebook learning. 
        We also propose to learn separate codebooks for line-of-sight and non-line-of-sight users, leveraging the geometric information provided by the digital twin.
        Simulation results demonstrate that the codebook learned from the digital twin can adapt to the environment geometry and user distribution, leading to high received signal-to-noise ratio performance.
        Moreover, we identify the ray-tracing accuracy as the most critical factor in digital twin fidelity that impacts the learned codebook performance.
    \end{abstract}

    \section{Introduction}
    Millimeter wave (mmWave) communication has emerged as a key technology in 5G and future-generation wireless systems. With its large available bandwidth, mmWave can deliver high data rates to support emerging data-intensive applications, such as collaborative artificial intelligence agents, autonomous vehicles, and immersive extended reality. To compensate for the high path loss, mmWave systems employ large antenna arrays that focus the signal and ensure sufficient received power at the user. The increasing number of antennas, however, makes the conventional fully-digital array architecture impractical due to the high hardware costs and power consumption of the mixed-signal circuits at the mmWave band. This motivates the development of analog-only and hybrid analog/digital architectures~\cite{Hur2013,Alkhateeb2014}, where the number of radio-frequency (RF) chains is much lower than the number of antennas. Since the channel can only be estimated through the limited number of the RF chains, acquiring accurate channel state information becomes challenging, which complicates the beamforming design. Hence, mmWave wireless systems require more efficient channel estimation and beamforming design approaches.

    In the literature, a common solution to the mmWave beamforming design problem is to pre-define a generic beamforming codebook consisting of a set of single-lobe beams that cover the entire angular domain~\cite{Giordani2019}. Specifically, the base station scans the codebook and selects the beam that maximizes the received signal power at the user. This classical approach, however, has several downsides. (i) It is a generic solution that does not account for the site-specific channel characteristics, such as environment geometry and user distribution. This can lead to a high beam training overhead, since the BS needs to scan all possible directions to find the optimal beam. (ii) The codebook design typically assumes a calibrated array and known geometry, whereas the calibration process for unknown or arbitrary array geometry is expensive in practice. To address these challenges, machine learning (ML) techniques have been utilized to learn the site-specific beamforming codebooks from partial channel information~\cite{Alrabeiah2022} or received signal power measurements~\cite{Zhang2022}. Nonetheless, the existing approaches require a large amount of data and iterations to train the model, which may not be practical and may lead to high training overhead.

    In this paper, we propose using site-specific digital twins~\cite{Alkhateeb2023} to reduce the data collection and learning burden associated with existing data-driven codebook learning approaches. The contributions of this paper are summarized as follows:
    \begin{itemize}
        \item We introduce a novel digital twin aided codebook learning framework for mmWave MIMO systems, leveraging site-specific digital twins to generate synthetic channel data for training codebooks. In particular, we propose learning separate codebooks for line-of-sight (LoS) and non-line-of-sight (NLoS) users by utilizing the geometric information provided by the digital twin.
        \item We develop a digital twin dataset to evaluate the efficacy of the proposed framework and assess the impact of digital twin fidelity on the learned codebook performance. Extensive simulations demonstrate that our proposed approach achieves superior performance compared to the classical discrete Fourier transform (DFT) codebook. Also, we identify ray tracing accuracy as the most critical factor affecting the codebook learning performance.
    \end{itemize}

    \begin{figure*}
        \centering
        \includegraphics[width=0.7\textwidth]{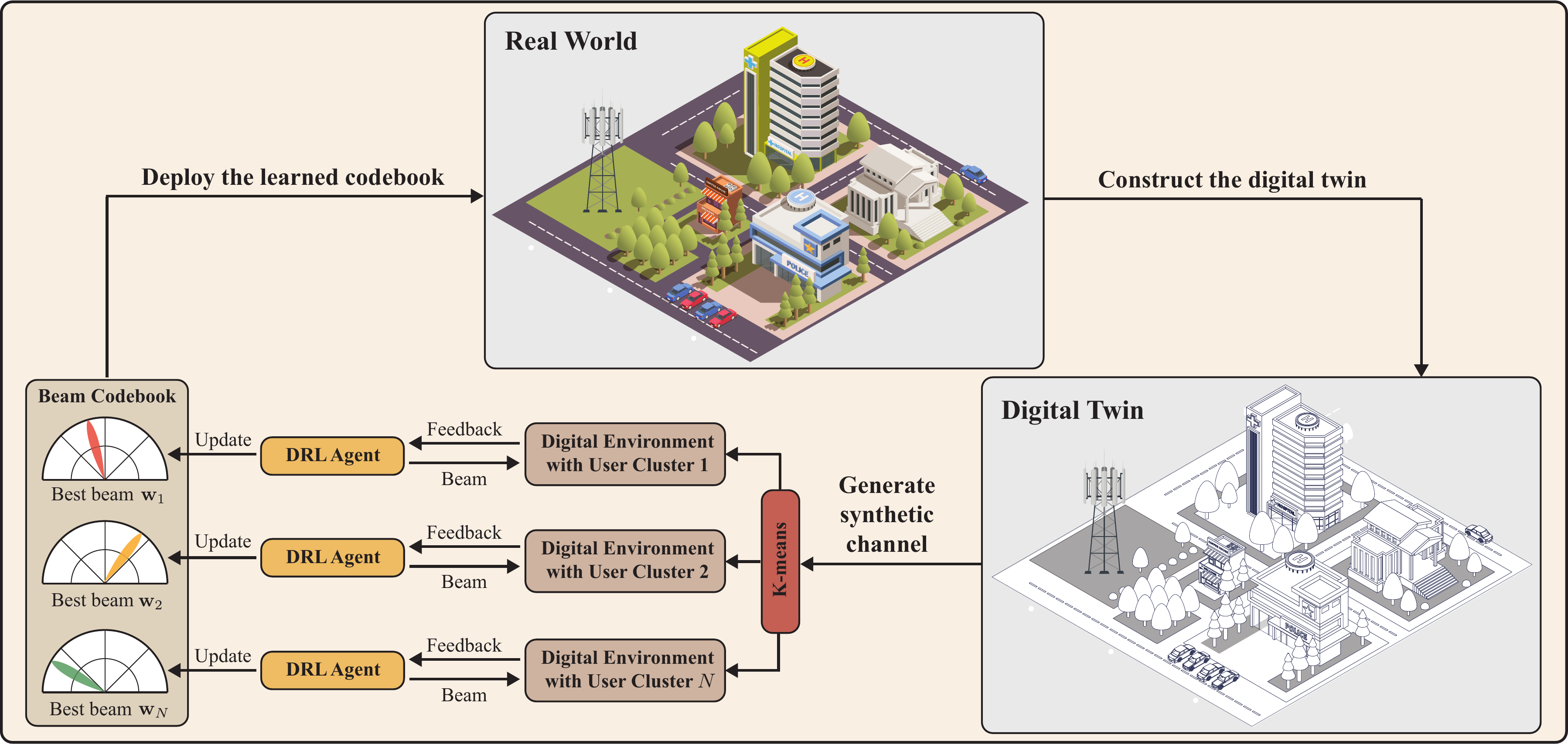}

        \caption{This diagram illustrates the proposed digital twin aided codebook learning framework. A digital replica of the real-world environment is constructed to generate synthetic channel information for codebook learning. The learned codebook can then be applied to real-world deployments.}
        \label{fig:system_model}
    \end{figure*}

    \section{System and Channel Models}
        We consider a mmWave MIMO system with a base station equipped with $M$ antennas and a single-antenna user. The base station employs an analog-only architecture with one RF chain and $r$-bit phase shifters. In the uplink transmission, the base station selects a combining vector $\bw \in \bbC^{M\times1}$ from a codebook $\bm{\cW}$ to receive the signal from the user. The codebook is assumed to consist of $N$ beamforming/combining vectors, each of which is given by
        \begin{equation} \label{eq:codebook}
            \bw = \frac{1}{\sqrt{M}}[e^{j\theta_{1}}, \ldots, e^{j\theta_{M}}]^T,
        \end{equation}
        where $\theta_{m} \in \bm{\Theta}$ is the phase shift of the $m^\mathrm{th}$ antenna element, and $\bm{\Theta}$ is the set of phase shift values that contains $2^r$ uniformly spaced values drawn from $(-\pi, \pi]$. The received signal at the base station is given by
        \begin{equation}
            y = \bw^H \bh x + \bw^H \bn,
        \end{equation}
        where the transmitted symbol $x \in \bbC$ satisfies the power constraint $\bbE[|x|^2] = P$, $\bh \in \bbC^{M\times1}$ is the uplink channel vector between the user and the base station, and $\bn \sim \cC\cN(0, \sigma^2 \bI)$ is the receive noise vector at the base station.
        For the channel vector $\bh$, we consider the geometric channel model, where the channel is composed of $L$ paths. Each path $l$ has a complex gain $\alpha_l$ and an angle of arrival at the base station $\phi_l$. The channel vector is given by
        \begin{equation}
            \bh = \sum_{l=1}^{L} \alpha_l \ba(\phi_l),
        \end{equation}
        where $\ba(\phi) = [1, e^{jkd\cos(\phi)}, \ldots, e^{jkd(M-1)\cos(\phi)}]^T$ is the array response vector with $d$ being the antenna spacing and $k = \frac{2\pi}{\lambda}$ being the wave number.

    \section{Problem Formulation}
        In this paper, we aim to design a site-specific codebook learning framework that can adapt to the deployment, including the environment geometry, user distribution, given hardware, etc. To formulate the problem, we first define the received SNR after combining at the base station, given by
        \begin{equation}
            \mathsf{SNR} = \frac{|\bw^H \bh|^2}{\|\bw\|_2^2} \rho,
        \end{equation}
        with $\rho = \frac{P}{\sigma^2}$. Also, we define the combining gain with the adopted receive beamformer $\bw$ as
        \begin{equation}
            g = |\bw^H \bh|^2.
        \end{equation}
        With a codebook $\bm{\cW}$, the optimal SNR can be achieved via the exhaustive search, i.e.,
        \begin{equation} \label{eq:opt_snr}
            \mathsf{SNR}^\star = \rho \max_{\bw \in \bm{\cW}} |\bw^H \bh|^2,
        \end{equation}
        where $\|\bw\|^2 = 1$ since the phase shifters are adopted to implement the analog beamforming. Then, the goal is to design the optimal codebook $\bm{\cW}_{\mathsf{opt}}$ that maximizes the SNR in \eqref{eq:opt_snr} averaged over all users. Let $\bm{\cH}$ be the set of the channel vectors for all users served by the base station, the problem can be formulated as
        \begin{align} \label{eq:opt_codebook}
            \bm{\cW}_{\mathsf{opt}} & = \argmax_{\bm{\cW}} \frac{1}{|\bm{\cH}|} \sum_{\bh \in \bm{\cH}} \biggl(\max_{\bw_n \in \bm{\cW}} |\bw_n^H \bh|^2\biggr) , \\ \nonumber
            \text{s.t.} & \quad w_{mn} = \frac{1}{\sqrt{M}} e^{j\theta_{mn}}, \, \theta_{mn} \in \bm{\Theta}, \, \forall m,n,
        \end{align}
        where $w_{mn} = [\bw_n]_m$ is the $m^\mathrm{th}$ element of the $n^\mathrm{th}$ beamforming/combining vector $\bw_n \in \bm{\cW}$, and $\bm{\Theta}$ is the set of $2^r$ possible phase shift values. The constraint in \eqref{eq:opt_codebook} ensures that the beamformer can only apply quantized phase shift to the received signal.
        The optimization problem in \eqref{eq:opt_codebook} is challenging to solve due to the non-convexity of the constraint. Additionally, accurate channel information is required to solve the problem, which can be difficult to obtain in practice because of hardware limitations. In the literature, prior art has proposed learning the codebook from partial channel information~\cite{Alrabeiah2022} or received signal power measurements~\cite{Zhang2022} using online machine learning. However, these approaches require a relatively large amount of iterations to converge, leading to high learning overhead since the iterations are performed over the air. To address these challenges, we propose leveraging site-specific digital twins to generate synthetic channels for codebook learning. The learned site-specific codebook can then be directly applied to real-world deployments, thereby significantly reducing the data collection and learning overhead. 

    \section{Proposed Solution}
    \subsection{Key Idea}
    In this section, we present the proposed digital twin aided codebook learning framework for mmWave MIMO systems, as shown in \figref{fig:system_model}. The key idea is to use a site-specific digital twin, which is a virtual replica of the real-world communication environment, to generate synthetic channels for data-driven codebook design. In the real world, a communication channel is determined by: (i) communication environment $\cE$, including positions, orientations, dynamics, shapes, and electromagnetic (EM) properties of the objects in the environment, (ii) wireless propagation law $\cG(\cdot)$, and (iii) hardware characteristics of the communication devices $\cD$. Thus, real-world communication channels $\bm{\cH}$ can be expressed as
    \begin{equation}
        \bm{\cH} = \cG(\cE, \cD).
    \end{equation}
    In the digital twin, the real-world communication environment is approximated by an EM 3D model $\widetilde{\cE}$. The signal propagation is simulated using an EM solver $\widetilde{\cG}(\cdot)$, e.g., a ray-tracing tool. Then, the synthetic channels $\widetilde{\bm{\cH}}$ can be generated as
    \begin{equation}
        \widetilde{\bm{\cH}} = \widetilde{\cG}(\widetilde{\cE}, \cD),
    \end{equation}
    where the hardware characteristics $\cD$ are assumed to be measured in controlled environments by the device manufacturer and are available to the digital twin.

    The digital twin not only provides a way to generate synthetic channel data, but also introduces new opportunities and challenges for codebook learning. In this work, we focus on the following two aspects:
    \begin{itemize}
        \item \textbf{Designing separate codebooks for LoS and NLoS users.} Digital twins can help classify users into LoS and NLoS categories based on the environment geometry. This classification allows for the design of separate codebooks tailored to the distinct channel characteristics of each user type, which can enhance overall system performance without increasing beam training overhead.
        \item \textbf{Assessing the impact of digital twin fidelity.} The accuracy of the digital twin model can affect the quality of the synthetic channel data. In our simulation, we investigate how different aspects of digital twin fidelity impact the performance of the learned codebook, providing insights into which elements of the digital twin are most critical for effective codebook learning.
    \end{itemize}
    To demonstrate the efficacy of digital twins, we adopt a reinforcement learning (RL) based codebook learning method~\cite{Zhang2022} as a case study. In the following, we present the details of the proposed digital twin aided codebook learning framework.

    \subsection{Digital Twin Aided Codebook Learning}
    The codebook design problem in \eqref{eq:opt_codebook} is essentially a combinatorial optimization problem, which has a huge yet finite search space. To reduce the difficulty of the problem, we decouple the codebook design into $N$ independent subproblems, where each subproblem corresponds to the design of a single beamforming/combining vector. The motivation for this decoupling is that the number of beamforming/combining vectors $N$ is typically much smaller than the number of users, and the users with similar channel characteristics can share the same beamforming/combining vector. The beamforming/combining vector design problem can be formulated as
    \begin{align} \label{eq:opt_beam}
        \bw_{\mathsf{opt}} & = \argmax_{\bw} \frac{1}{|\widetilde{\bm{\cH}}_s|} \sum_{\bh \in \widetilde{\bm{\cH}}_s} |\bw^H \bh|^2 , \\ \nonumber
        \text{s.t.} & \quad w_{m} = \frac{1}{\sqrt{M}} e^{j\theta_{m}}, \, \theta_{m} \in \bm{\Theta}, \, \forall m,
    \end{align}
    where $w_m$ is the $m^\mathrm{th}$ element of the beamforming/combining vector $\bw$, and $\widetilde{\bm{\cH}}_s$ is the set of synthetic channel vectors from the digital twin. Next, we present the clustering approach to group the synthetic channels.

    \textbf{User clustering.} By clustering the synthetic channels, we can group the user positions with similar channel characteristics and share the same beamforming/combining vector. We adopt an SNR-based clustering approach~\cite{Zhang2022}, where the user positions are clustered based on the measurements of the received combining gain at the base station. The clustering algorithm starts by constructing a set of sensing beams, which are randomly sampled from the feasible phase shift values. Let $\cF = \{\bff_1, \ldots, \bff_S\}$ be the set of $S$ sensing beams,where $\bff_s \in \bbC^{M\times1}$, $\forall s\in\{1,\ldots,S\}$. Also, we assume there are $K$ user positions in the digital twin, and the synthetic channel vectors are denoted by $\widetilde{\bm{\cH}} = \{\widetilde{\bh}_1, \ldots, \widetilde{\bh}_K\}$. The collected received combining gain for the $K$ user positions with the sensing beams can form a sensing matrix $\bP$, given by
    \begin{equation}
        \bP = \begin{bmatrix}
            |\bff_1^H \widetilde{\bh}_1|^2 & \cdots & |\bff_1^H \widetilde{\bh}_K|^2 \\
            \vdots & \ddots & \vdots \\
            |\bff_S^H \widetilde{\bh}_1|^2 & \cdots & |\bff_S^H \widetilde{\bh}_K|^2 \\
        \end{bmatrix}.
    \end{equation}
    The columns of this matrix capture the channel characteristics of each user position. A clustering algorithm, such as the K-means algorithm, can then be applied to the columns of $\bP$ to partition the synthetic channels into $N$ clusters, i.e., $\widetilde{\bm{\cH}} = \widetilde{\bm{\cH}}_1\cup \ldots \cup \widetilde{\bm{\cH}}_N$, where $\widetilde{\bm{\cH}}_k \cap \widetilde{\bm{\cH}}_l = \emptyset$ for $k\neq l$. The beamforming/combining vector design problem in \eqref{eq:opt_beam} can then be solved for each cluster independently.

    \textbf{Reinforcement learning setup.}
    We adopt RL to solve the beamforming/combining vector design problem in \eqref{eq:opt_beam}. The components of the RL framework are defined as follows:
    \begin{itemize}
        \item \textbf{State:} The state is defined as the phases of the phase shifters, and $\bs_t$ denotes the state at the $t^\mathrm{th}$ iteration, i.e., $\bs_t = [\theta_{1}, \ldots, \theta_{M}]^T$. The actual beamforming/combining vector can be obtained by applying \eqref{eq:codebook} to the phase vector.
        \item \textbf{Action:} The action $\ba_t$ is defined as the element-wise changes to the phase shift values in $\bs_t$, which means the phase shifter select a new phase value from the set of possible phase shift values $\bm{\Theta}$. The action is specified as the next state, i.e., $\bs_{t+1} = \ba_t$.
        \item \textbf{Reward:} The reward $r_t \in \{-1, +1\}$ is determined by comparing the current combining gain $g_t$ with two values: (i) An adaptive threshold $\beta_t$, and (ii) the previous received combining gain $g_{t-1}$. The reward is computed based on the following rules: $r_t = +1$ if $g_t > \beta_t$; $r_t = +1$ if $g_t \leq \beta_t$ and $g_t > g_{t-1}$; otherwise, $r_t = -1$.
    \end{itemize}
    The threshold $\beta_t$ starts at zero and is updated to the maximum combining gain observed so far. When $\beta_t$ increases, the corresponding beamforming vector is stored in the codebook. After training, the codebook contains the beams achieving the maximum gain for each cluster.

    \textbf{Deep reinforcement learning with digital twin.}
    We adopt the deep deterministic policy gradient (DDPG) algorithm~\cite{Lillicrap2015} to learn combining vectors within a digital twin environment. DDPG uses two neural networks: An actor network ($\mu_\varphi$) that predicts an action from a given state, and a critic network ($Q_\omega$) that estimates the Q-value (expected future reward) for a state-action pair. The actor network outputs a proto-action $\widehat{\ba}_t = \mu_\varphi(\bs_t)$. This proto-action is then quantized to produce a feasible action $\ba_t$ via an element-wise quantization operation, which is defined as
    \begin{equation}
        [\ba_t]_m = \argmin_{\theta \in \bm{\Theta}} |\theta - [\widehat{\ba}_t]_m|, \, \forall m.
    \end{equation}
    Then, we can convert the action $\ba_t$ to the beamforming/combining vector $\widetilde{\bw}_t$ by applying \eqref{eq:codebook}. This beam can be used to interact with the digital twin to obtain the received combining gain $\bar{g}_t$ averaged over all synthetic channels $\widetilde{\bm{\cH}}_n$ in the cluster, i.e.,
    \begin{equation}
        \bar{g}_t = \frac{1}{|\widetilde{\bm{\cH}}_n|} \sum_{\bh \in \widetilde{\bm{\cH}}_n} |\widetilde{\bw}_t^H \bh|^2.
    \end{equation}
    Depending on the received combining gain $\bar{g}_t$, the reward $r_t$ is computed as described above. The new state is updated as $\bs_{t+1} = \ba_t$. The transition tuple $\{\bs_t, \ba_t, r_t, \bs_{t+1}\}$ is then stored in the replay buffer. The actor and critic networks are trained using a mini-batch of $N$ transition tuples sampled from the replay buffer. The actor's objective is to maximize the Q-value predicted by the critic. The update uses a gradient ascent step:
    \begin{align}
        \nabla_\varphi J(\varphi) &\approx \frac{1}{N} \sum_{i=1}^N \nabla_\ba Q_\psi(\bs_i, \ba)|_{\ba = \mu_\varphi(\bs_i)} \nabla_\varphi \mu_\varphi(\bs_i).
    \end{align}
    The critic network is trained to minimize the mean squared error between the predicted Q-value and the target Q-value. The loss function for the critic network is given by
    \begin{align}
        L(\omega) = \frac{1}{N} \sum_{i=1}^N \biggl( y_i - Q_\omega(\bs_i, \ba_i) \biggr)^2,
    \end{align}
    where $y_i = r_i + \gamma \, Q_{\psi_{\rm{targ}}}(\bs_{i+1}^\prime, \mu_{\varphi_{\rm{targ}}}(\bs_{i+1}^\prime))$ is the target Q-value, and $\gamma \in [0, 1]$ is the discount factor. To stabilize training, target networks $\psi_{\rm{targ}}$ and $\varphi_{\rm{targ}}$ are used, which are soft-updated from the main networks with a parameter $\tau \in (0, 1]$, as described below.
    \begin{align}
        \psi_{\rm{targ}} \leftarrow (1 - \tau) \psi_{\rm{targ}} + \tau \psi, \\
        \varphi_{\rm{targ}} \leftarrow (1 - \tau) \varphi_{\rm{targ}} + \tau \varphi.
    \end{align}
    Additionally, an Ornstein-Uhlenbeck (OU) process~\cite{Uhlenbeck1930} is used to add noise to the action for exploration during training.

    \subsection{Separate Codebooks for LoS and NLoS Users} 
    In this subsection, we present the proposed strategy of learning separate site-specific codebooks for LoS and NLoS users. This approach is motivated by the observation that these two user types have distinct channel characteristics due to differences in multi-path propagation. By using separate codebooks, we can tailor beamforming strategies to better suit each type of user, which can improve overall system performance without increasing the beam training overhead. The prerequisite for this method is the ability to classify users as either LoS or NLoS. This can be done by analyzing signal strength or using environmental information from a digital twin. A digital twin makes this task especially straightforward by checking for a direct path between the base station and the user, without needing highly precise location information unless the user is at the border of LoS and NLoS regions. Once users are classified, we can apply the digital twin aided codebook learning framework separately to each group. This involves generating synthetic channel data for both LoS and NLoS users using the digital twin and training distinct DRL agents. These agents learn an optimal codebook for each category. The learned codebooks can then be deployed in real-world systems, with the base station selecting the appropriate codebook based on the user's classification. This approach leads to a more efficient codebook design that accounts for the unique characteristics of LoS and NLoS channels.

    \section{Simulation Results}
    \subsection{Simulation Setup}

    \begin{figure}[t]
        \centering
        \includegraphics[width=0.275\textwidth]{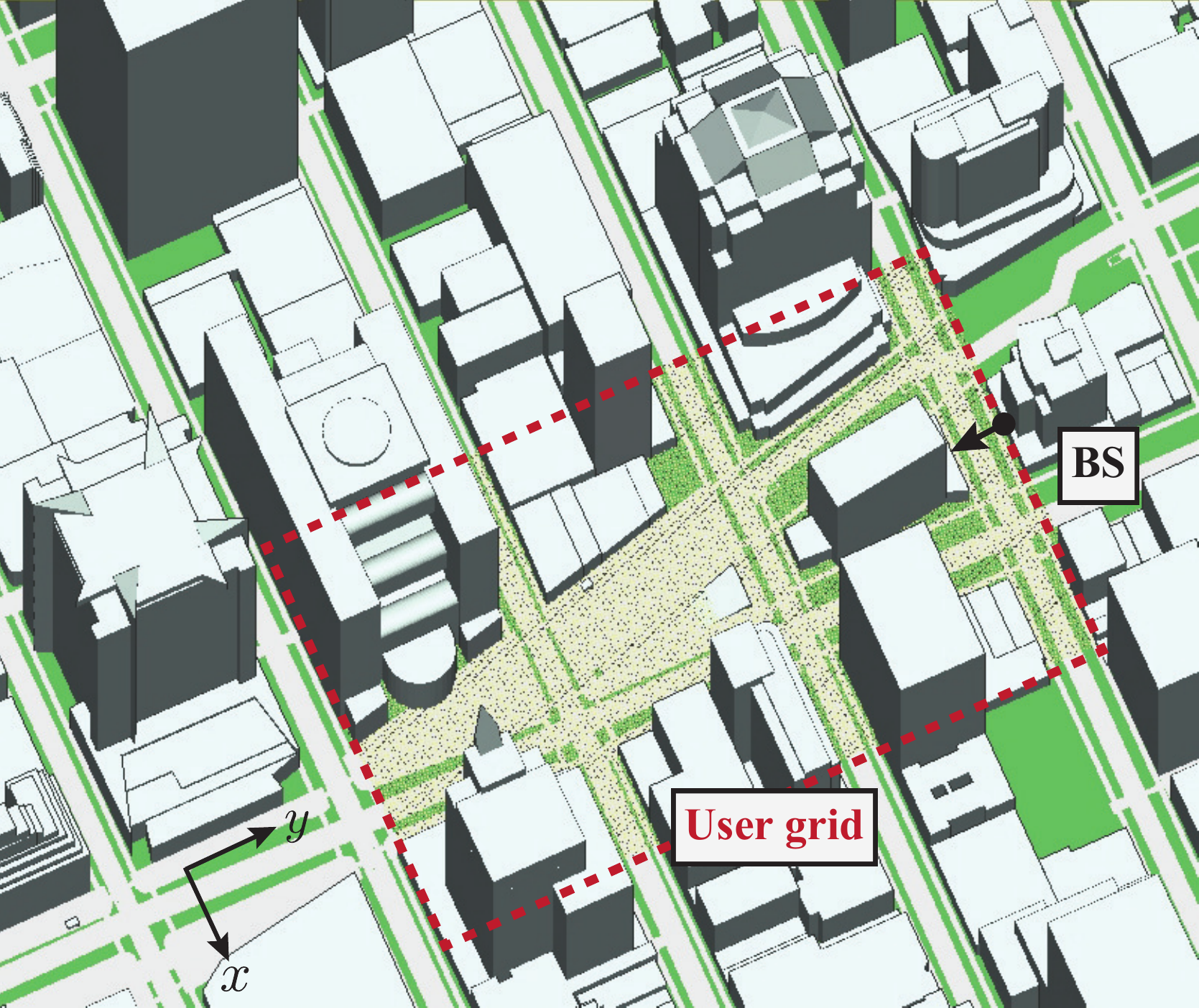}
        \caption{The figure shows the adopted target scenario, which is built based on Times Square in Manhattan. The base station is positioned at a building, and the user grid is highlighted by the red box.}
        \label{fig:times_square}
    \end{figure}

    \textbf{Target scenario.} The target scenario represents the "real-world" deployment in our simulation. As shown in \figref{fig:times_square}, we adopt an urban scenario modeled after Times Square in Manhattan. This scenario includes a BS, a service area, buildings, terrain, and streets. The BS is assumed to employ an $4 \times 8$ uniform planar array and an analog-only architecture. It is placed at a height of $20$ meters and oriented towards the negative $y$-axis. The UEs are located in the service area at a height of $1.5$ meters. The service area measures $145$ meters by $238$ meters and is discretized into a user grid with a spacing of $0.5$ meters. For EM materials, the ITU concrete, ITU wet earth, and asphalt are used to model the buildings, terrain, and streets, respectively. For the ray tracing parameters, we assume the propagation paths between the BS and UE are searched up to the $4^{\rm{th}}$ order reflection.

    \textbf{Digital twin scenario.} 
    Building a perfect digital twin is challenging, as its fidelity can be impacted by several factors. We analyze the impact of fidelity aspects including the 3D geometry model, EM material, and ray tracing~\cite{Luo2025}. Specifically, we assume that a 3D geometry model is constructed based on LiDAR point clouds with a sampling density of $0.5$ points$/\textrm{m}^2$. This can be simulated by sampling the surfaces of the objects in the target scenario. The sampled points are then used to reconstruct the 3D geometry model using the Poisson surface reconstruction algorithm~\cite{Kazhdan2006}. All objects are assumed to be made of concrete, and the ray tracing is limited to first-order reflections. The digital twin and target scenario are assumed to have identical hardware characteristics.

    \textbf{Dataset generation.}
    We adopt Wireless Insite~\cite{Remcom} to conduct ray tracing simulations in both the target and digital twin scenarios. The ray tracing simulations operates at a carrier frequency of $28$ GHz. For each path, the complex gain $\alpha_l$ and the angle of arrival $\phi_{l}$ are obtained. Then, we use the DeepMIMO generator~\cite{Alkhateeb2019} to construct the channel. The target scenario dataset includes 14,559 user positions (4,108 LoS and 10,451 NLoS). The digital twin scenario dataset has 6,189 user positions (4,163 LoS and 2,026 NLoS). The digital twin has fewer user positions because of the limited NLoS coverage due to low ray tracing accuracy.

    \textbf{Deep learning architecture.}
    The actor and critic networks are implemented as fully-connected neural networks. The input to the actor network is the state, i.e., the phases of the phase shifters. The actor network has two hidden layers with $16M$ neurons followed by ReLU activation. The output of the actor network is the predicted action. Thus, the output layer has $M$ neurons followed by hyperbolic tangent (tanh) activation scaled by $\pi$, making the output in the range of $(-\pi, \pi]$. The critic network has the same structure as the actor network, except that the input is the concatenation of the state and action, and the output is the predicted Q-value, which is a real scalar.

    \begin{figure}[t]
        \centering
        
        \subfigure[LoS users]{\label{fig:snr_cdf_los_separate_cb}\includegraphics[width=0.24\textwidth]{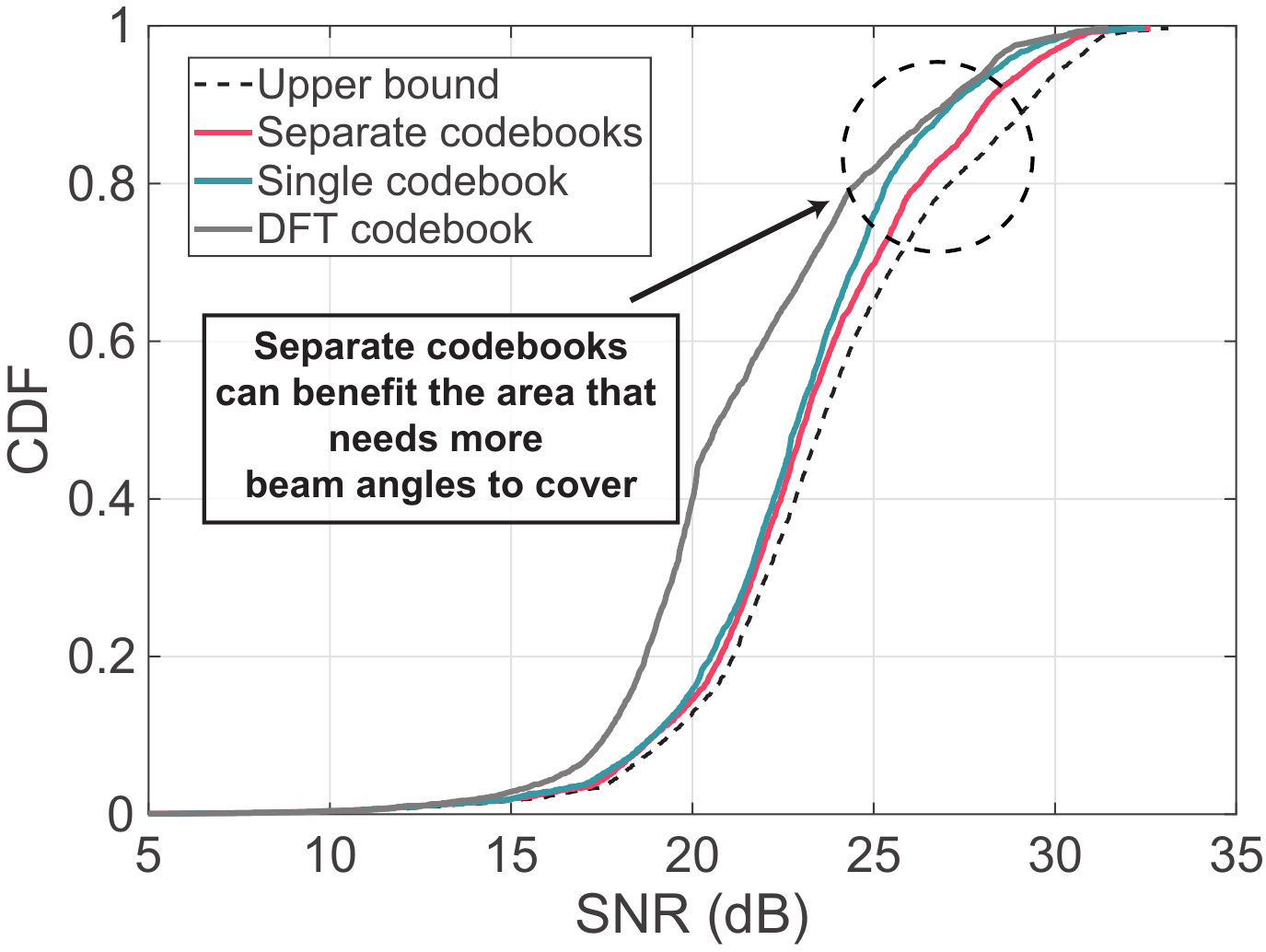}}
        \subfigure[NLoS users]{\label{fig:snr_cdf_nlos_separate_cb}\includegraphics[width=0.24\textwidth]{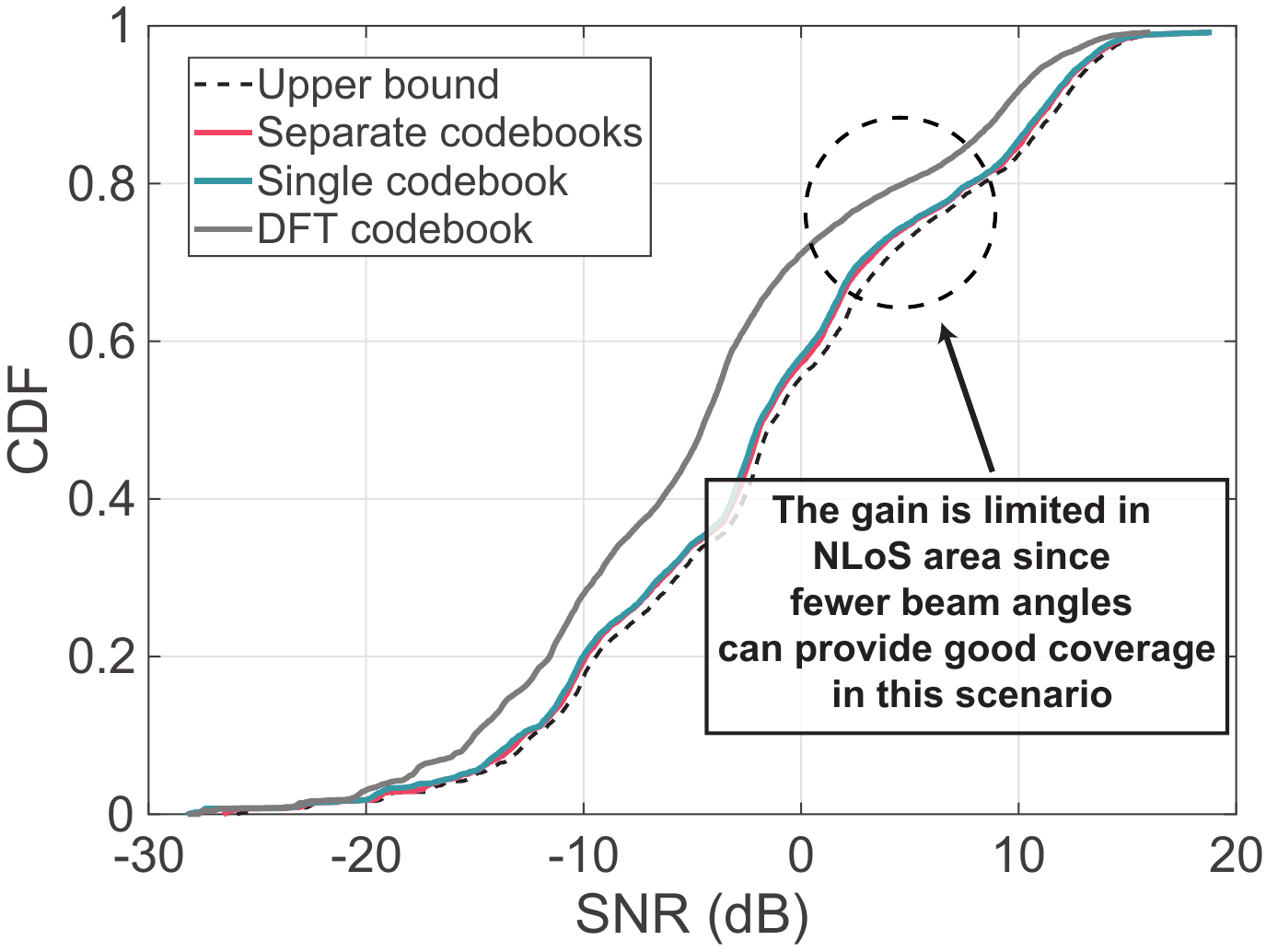}}

        \caption{This figure shows the CDF of the received SNR in the target scenario for using separate codebooks for LoS and NLoS users compared to using a single codebook. The results indicate that employing separate codebooks enhances performance in the regions that require more beam angles to cover.}
        \label{fig:snr_cdf_separate_cb}
    \end{figure}

    \begin{figure}[t]
        \centering

        \subfigure[LoS users]{\label{fig:snr_cdf_los}\includegraphics[width=0.24\textwidth]{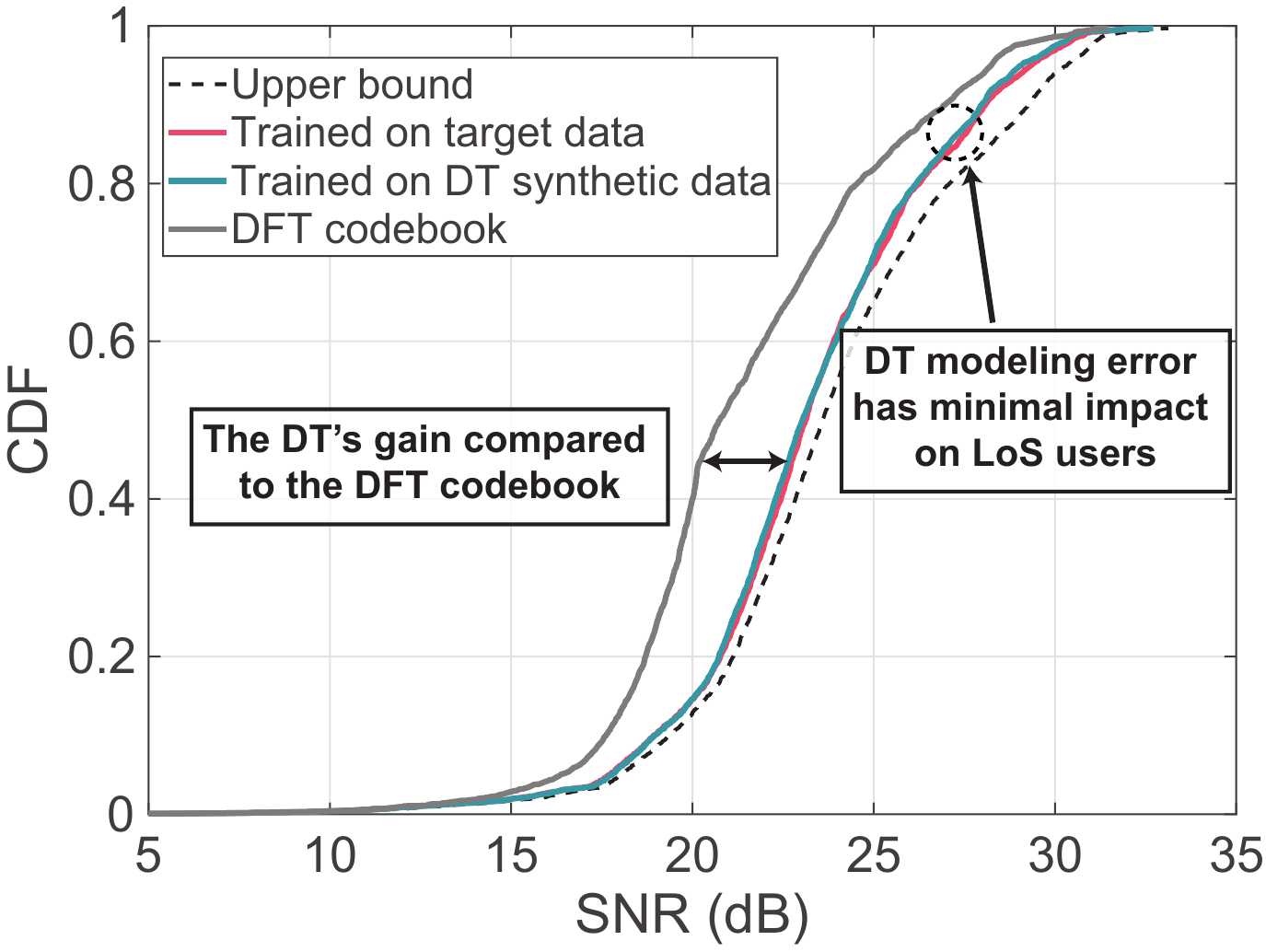}}
        \subfigure[NLoS users]{\label{fig:snr_cdf_nlos}\includegraphics[width=0.24\textwidth]{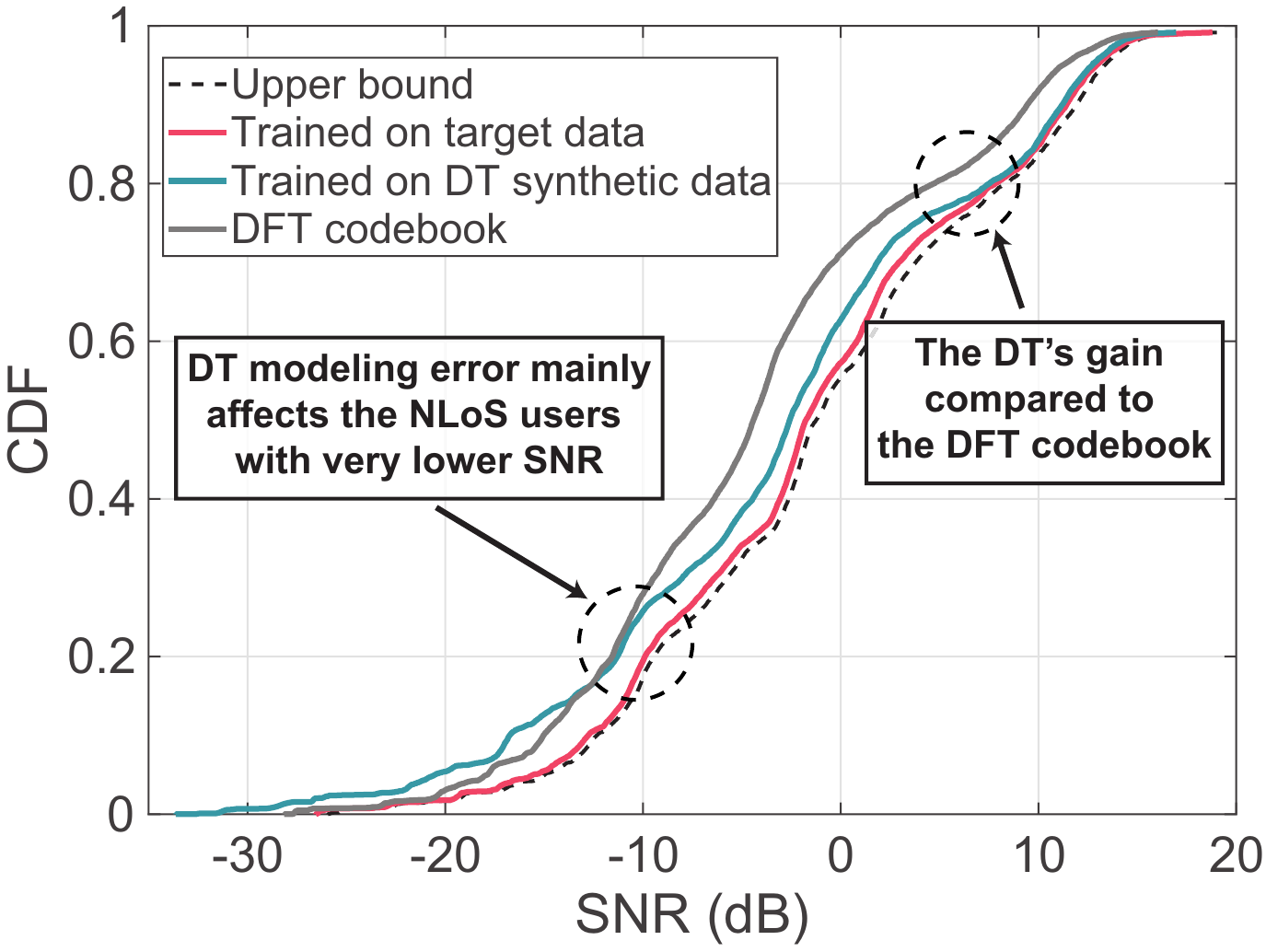}}

        \caption{This figure shows the CDF of the received SNR for the target scenario, the digital twin (DT) scenario, and the DFT beams. The results show that the codebook learned from the digital twin can achieve better performance than the DFT codebook in both LoS and NLoS regions.}
        \label{fig:snr_cdf}
    \end{figure}

    \subsection{Performance Evaluation}
    The received SNR at the user is used as the main evaluation metric. To compute it, we assume the equivalent isotropic radiated power of the base station is $15$ dBm, and the noise figure of the user equipment is $5$ dB.

    \textbf{What is the benefit of using separate codebooks for LoS and NLoS users?}
    In \figref{fig:snr_cdf_separate_cb}, we present the cumulative distribution function (CDF) of the received SNR in the target scenario when using separate codebooks for LoS and NLoS users compared to using a single codebook for all users. The results show that employing separate codebooks enhances performance in regions that require more beam angles to cover, such as the LoS region in the adopted scenario. For the NLoS region, since the signal can only reach the user through the two narrow streets, the channel characteristics are more similar among different user positions. Thus, using separate codebooks provides a smaller performance gain in the NLoS region. Overall, this improvement is attributed to the geometric information provided by the digital twin, which helps identify the distinct channel characteristics of LoS and NLoS users.
    
\begin{figure}[t]
        \centering
        
        \subfigure[LoS users]{\label{fig:SA_los}\includegraphics[width=0.322\textwidth]{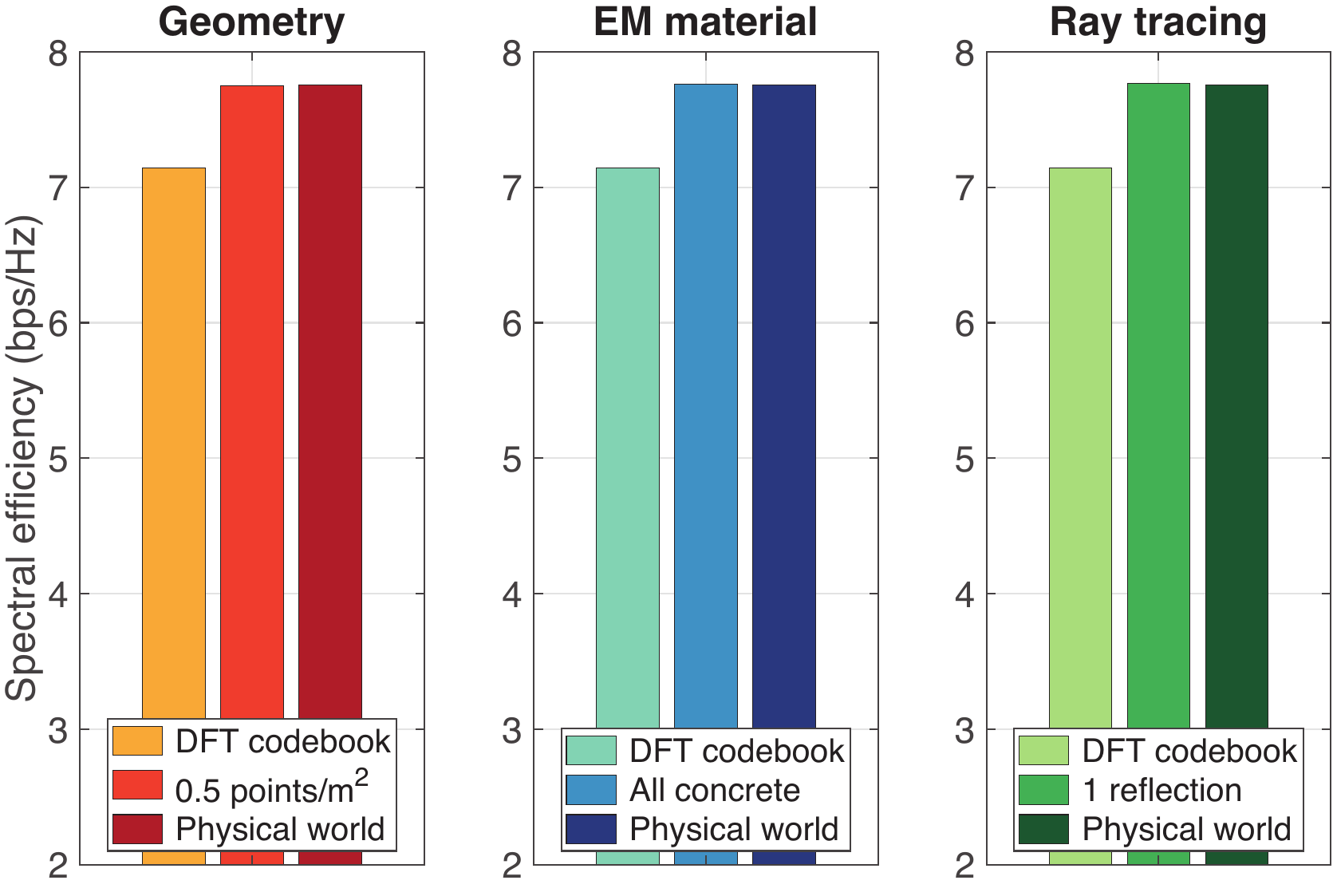}}
        \subfigure[NLoS users]{\label{fig:SA_nlos}\includegraphics[width=0.322\textwidth]{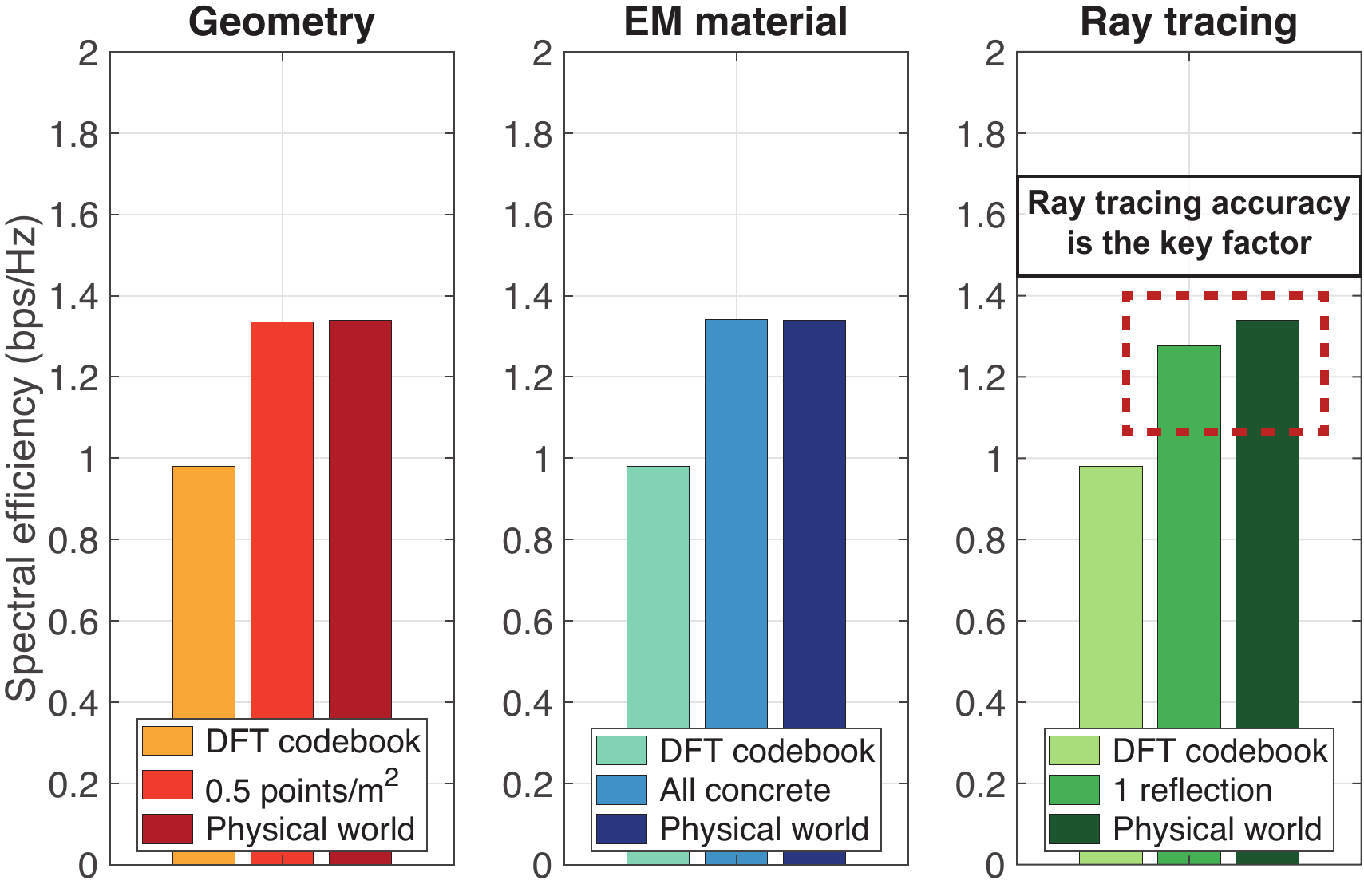}}

        \caption{This figure presents the sensitivity analysis of the proposed approach with respect to the 3D geometry, EM material, and ray tracing. The results show that the ray tracing accuracy has a significant impact on the performance, while the geometry and EM material have a relatively minor effect.}
        \label{fig:sensitivity_analysis}
    \end{figure}

    \textbf{Can the codebook learned from the digital twin work effectively and what is the key fidelity factor?}
    In \figref{fig:snr_cdf}, we present the CDF of the received SNR and compare the performance of the proposed approach with the following benchmarks: (i) equal-gain combining, where the channel information is assumed to be perfectly known, (ii) codebook learning with the target scenario, and (iii) DFT codebook. All methods are evaluated in the target scenario. For both LoS and NLoS users, the proposed digital twin based approach significantly outperforms the DFT codebook. This demonstrates that the synthetic channel data generated by the digital twin can be used to effectively learn site-specific codebooks. Due to the imperfection in digital twin modeling, the proposed approach experiences performance loss compared to the target scenario, especially for the NLoS users with lower SNR. To investigate the cause of this performance loss, we conducted a sensitivity analysis to examine the impact of each factor on performance. This was done by adjusting one parameter of digital twin fidelity at a time while keeping the other aspects fixed. In \figref{fig:sensitivity_analysis}, we can observe the codebook learning performance is sensitive to the ray-tracing accuracy, while the geometry and EM material has a relatively minor impact on the performance. This is because the ray tracing determines if the propagation paths between the BS and UE exist, and reducing the number of reflections can significantly degrade the coverage of the NLoS region and affect the channel characteristics.
    
    \begin{figure}[t]
        \centering
        
        \subfigure[Learned beam of DT scenario]{\label{fig:beam_DT}\includegraphics[width=0.23\textwidth]{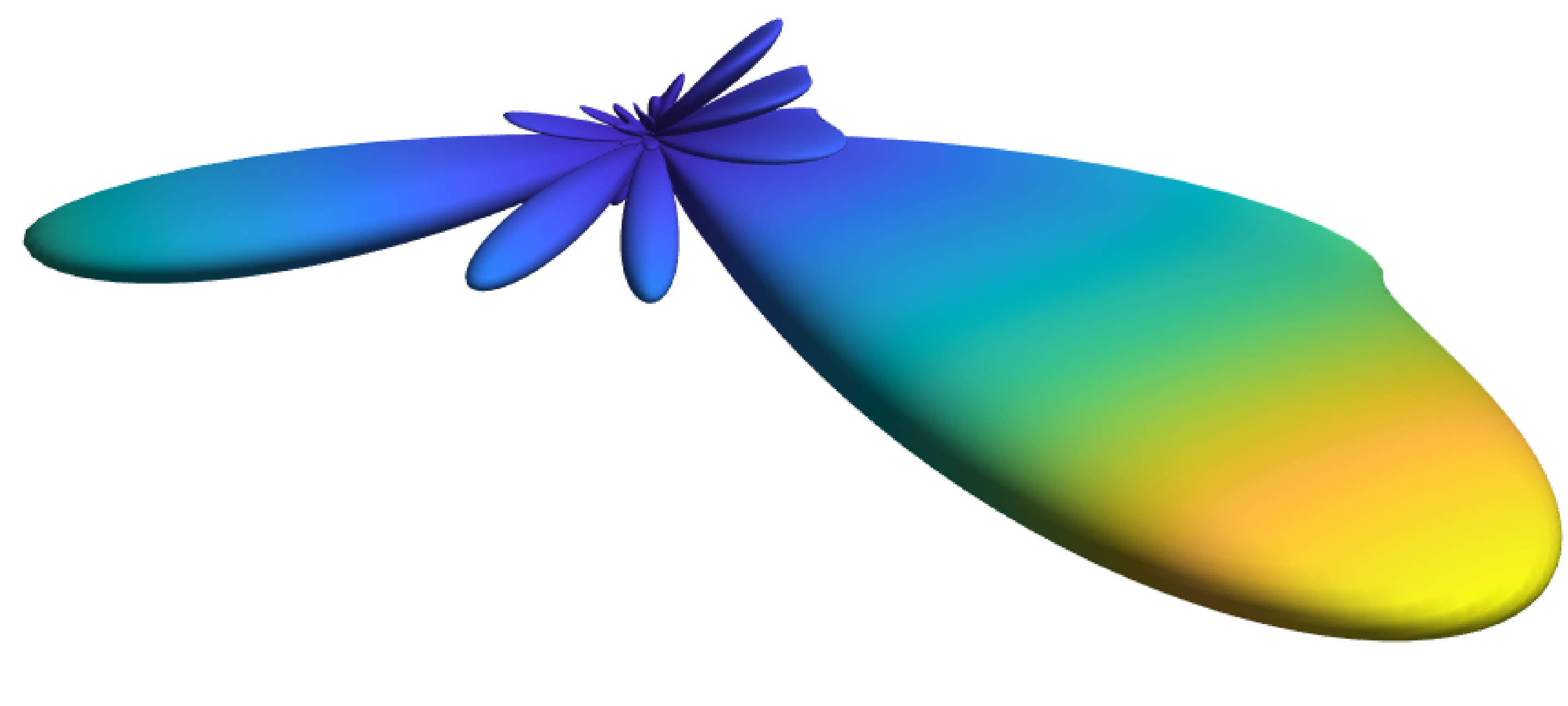}}
        \subfigure[DFT beam]{\label{fig:beam_DFT}\includegraphics[width=0.23\textwidth]{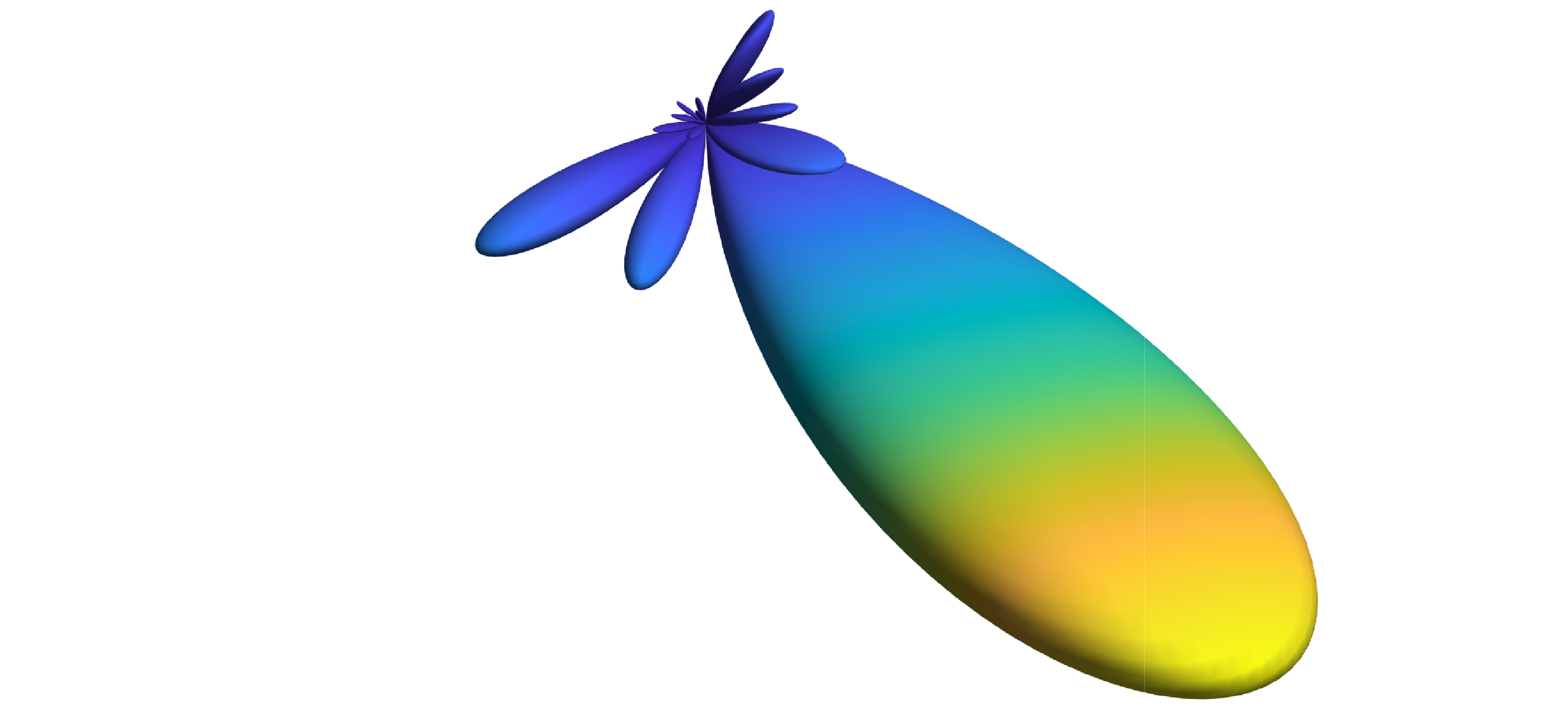}}
        \subfigure[SNR map of DT scenario]{\label{fig:snr_DT}\includegraphics[width=0.23\textwidth]{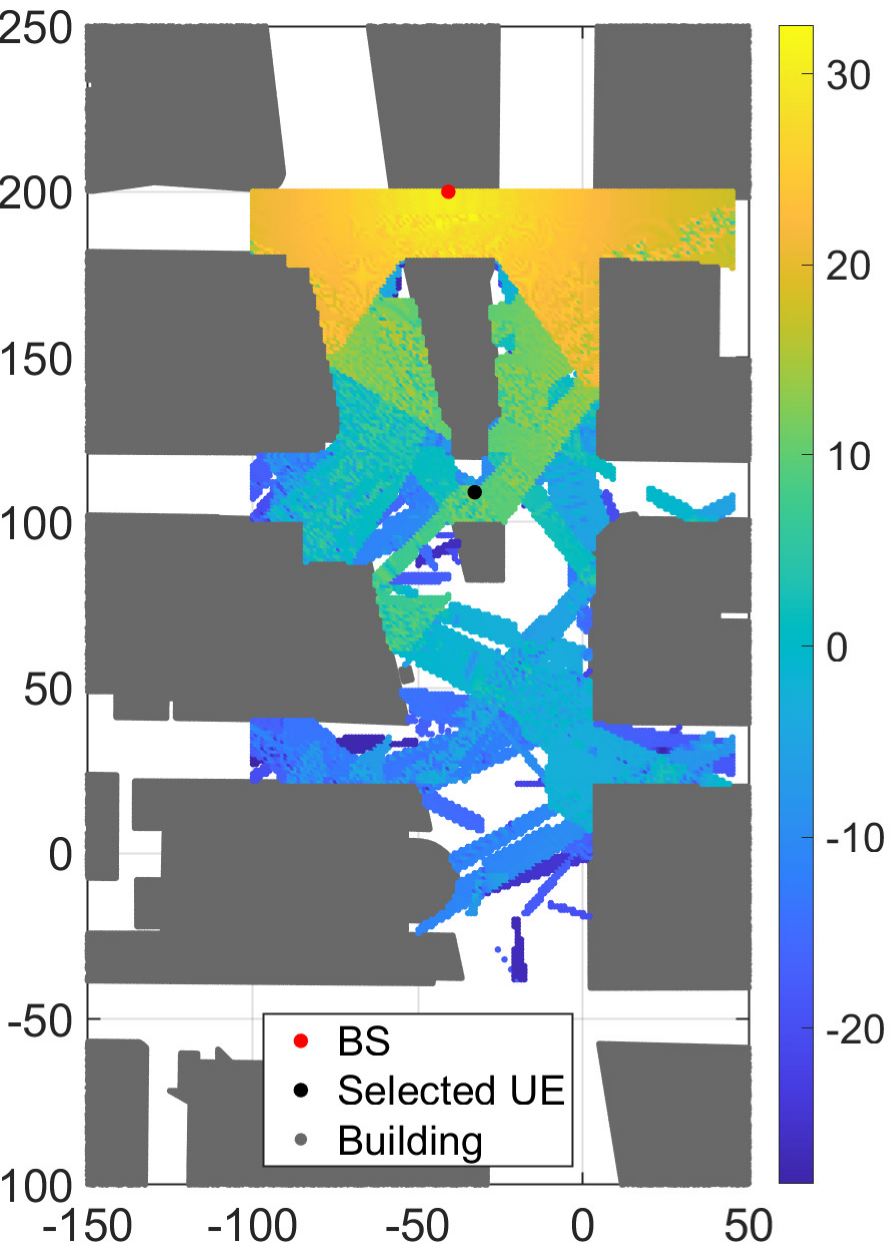}}
        \subfigure[SNR map of DFT beams]{\label{fig:snr_DFT}\includegraphics[width=0.23\textwidth]{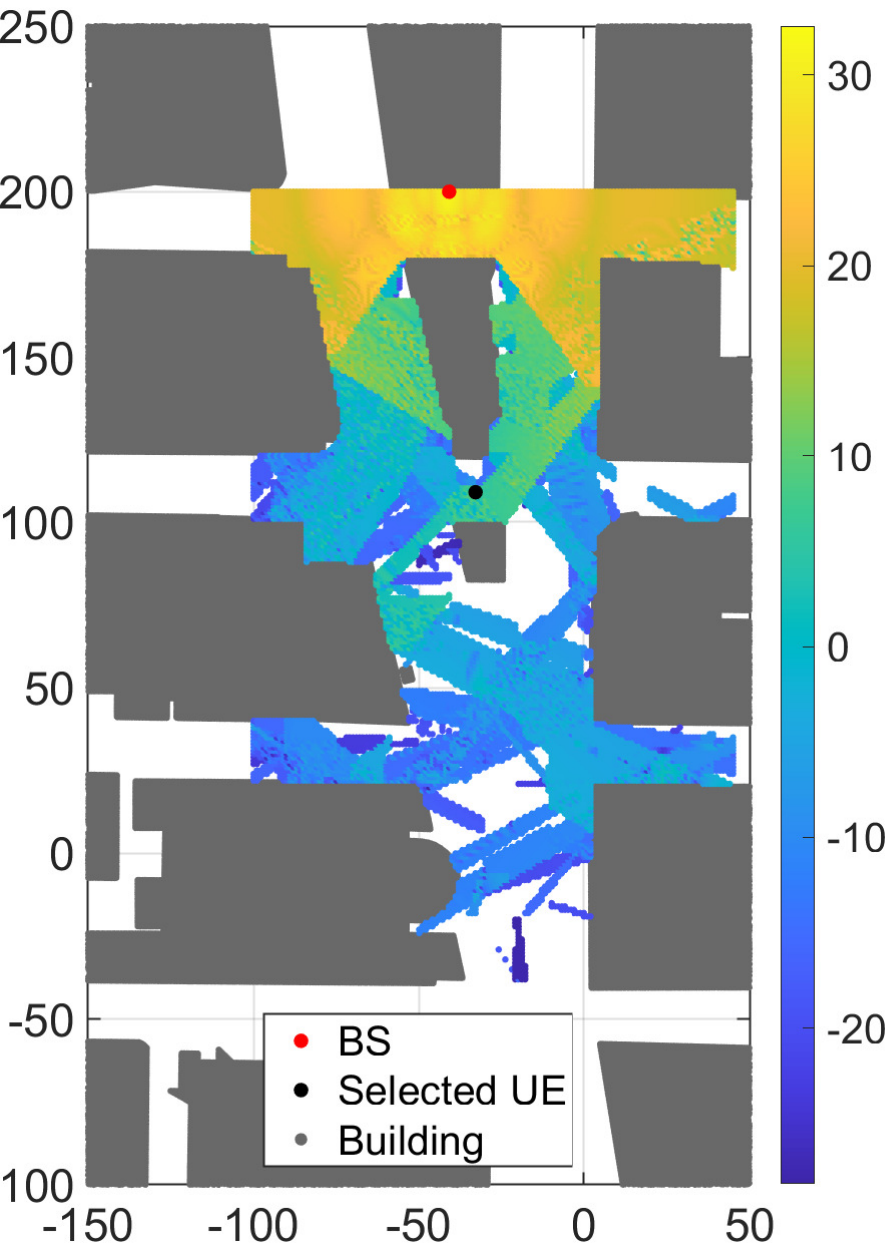}}

        \caption{This figure shows the beam patterns that achieve the maximum received SNR at a specific user position and SNR maps for the digital twin (DT) scenario and DFT beams. The learned codebook can adapt to the environment, achieving better performance in both LoS and NLoS regions.}
        \label{fig:snr_map}
    \end{figure}

    \textbf{Can the learned codebook adapt to the environment geometry and user distribution?}
    In \figref{fig:snr_map}, we present the learned beamforming patterns and plot the SNR values on the 2D layout of the target scenario. The presented beam patterns correspond to the maximum SNR achieved at a specific user position in the service area. We can observe that the DFT codebook suffers a performance degradation in both LoS and NLoS regions. This because the DFT codebook is designed for general purposes and does not take site-specific channel characteristics into account. In contrast, the proposed codebook learning approach can adapt to the environment geometry and user distribution, and the learned beams can effectively capture the promising directions of the real-world channels. This improvement stems from the fact that the learned beams offer greater flexibility in shape and angle compared to DFT beams, which allows them to better align with the actual channel conditions in both LoS and NLoS regions.

    \section{Conclusion}
    In this paper, we propose a digital twin aided codebook learning framework for mmWave MIMO systems. The proposed solution leverages site-specific digital twins to generate synthetic channel information for codebook learning, thereby reducing the data collection and learning overhead associated with existing ML-based codebook learning approaches. Using the digital twin, we propose learning separate codebooks for LoS and NLoS users by utilizing the geometric information provided by the digital twin. Simulation results demonstrate that the proposed approach achieves high SNR performance in the target scenario, outperforming the classical DFT codebook.


\end{document}

%% file: input.tex
\usepackage{amsfonts}
\usepackage{times}
\usepackage{graphicx}
\usepackage{latexsym}
\usepackage{dsfont}
\usepackage{amssymb}
\usepackage{amsmath}
\usepackage{cite}
\usepackage{verbatim}

\newcommand{\figref}[1]{{Fig.}~\ref{#1}}

\def\bb0{{\mathbb{0}}}

\def\ba{{\mathbf{a}}}
\def\bb{{\mathbf{b}}}

\def\bff{{\mathbf{f}}}

\def\bh{{\mathbf{h}}}

\def\bm{{\mathbf{m}}}
\def\bn{{\mathbf{n}}}

\def\bs{{\mathbf{s}}}

\def\bw{{\mathbf{w}}}

\def\b0{{\mathbf{0}}}

\def\bI{{\mathbf{I}}}

\def\bP{{\mathbf{P}}}

\def\bbC{{\mathbb{C}}}

\def\bbE{{\mathbb{E}}}

\def\cC{\mathcal{C}}
\def\cD{\mathcal{D}}
\def\cE{\mathcal{E}}
\def\cF{\mathcal{F}}
\def\cG{\mathcal{G}}
\def\cH{\mathcal{H}}

\def\cN{\mathcal{N}}

\def\cW{\mathcal{W}}

\def\sf0{{\mathsf{0}}}









%% file: DT_CB_Learning.bbl
\begin{thebibliography}{10}
\providecommand{\url}[1]{#1}
\csname url@samestyle\endcsname
\providecommand{\newblock}{\relax}
\providecommand{\bibinfo}[2]{#2}
\providecommand{\BIBentrySTDinterwordspacing}{\spaceskip=0pt\relax}
\providecommand{\BIBentryALTinterwordstretchfactor}{4}
\providecommand{\BIBentryALTinterwordspacing}{\spaceskip=\fontdimen2\font plus
\BIBentryALTinterwordstretchfactor\fontdimen3\font minus \fontdimen4\font\relax}
\providecommand{\BIBforeignlanguage}[2]{{%
\expandafter\ifx\csname l@#1\endcsname\relax
\typeout{** WARNING: IEEEtran.bst: No hyphenation pattern has been}%
\typeout{** loaded for the language `#1'. Using the pattern for}%
\typeout{** the default language instead.}%
\else
\language=\csname l@#1\endcsname
\fi
#2}}
\providecommand{\BIBdecl}{\relax}
\BIBdecl

\bibitem{Hur2013}
S.~Hur \emph{et~al.}, ``{Millimeter Wave Beamforming for Wireless Backhaul and Access in Small Cell Networks},'' \emph{IEEE Transactions on Communications}, vol.~61, no.~10, pp. 4391--4403, 2013.

\bibitem{Alkhateeb2014}
A.~Alkhateeb \emph{et~al.}, ``{Channel Estimation and Hybrid Precoding for Millimeter Wave Cellular Systems},'' \emph{IEEE Journal of Selected Topics in Signal Processing}, vol.~8, no.~5, pp. 831--846, 2014.

\bibitem{Giordani2019}
M.~Giordani \emph{et~al.}, ``{A Tutorial on Beam Management for 3GPP NR at mmWave Frequencies},'' \emph{IEEE Communications Surveys \& Tutorials}, vol.~21, no.~1, pp. 173--196, 2019.

\bibitem{Alrabeiah2022}
M.~Alrabeiah \emph{et~al.}, ``{Neural Networks Based Beam Codebooks: Learning mmWave Massive MIMO Beams That Adapt to Deployment and Hardware},'' \emph{IEEE Transactions on Communications}, vol.~70, no.~6, pp. 3818--3833, 2022.

\bibitem{Zhang2022}
Y.~Zhang \emph{et~al.}, ``{Reinforcement Learning of Beam Codebooks in Millimeter Wave and Terahertz MIMO Systems},'' \emph{IEEE Transactions on Communications}, vol.~70, no.~2, pp. 904--919, 2022.

\bibitem{Alkhateeb2023}
A.~Alkhateeb \emph{et~al.}, ``{Real-Time Digital Twins: Vision and Research Directions for 6G and Beyond},'' \emph{IEEE Communications Magazine}, vol.~61, no.~11, pp. 128--134, 2023.

\bibitem{Lillicrap2015}
T.~P. Lillicrap \emph{et~al.}, ``{Continuous Control with Deep Reinforcement Learning},'' \emph{arXiv preprint arXiv:1509.02971}, 2015.

\bibitem{Uhlenbeck1930}
G.~E. Uhlenbeck and L.~S. Ornstein, ``{On the Theory of the Brownian Motion},'' \emph{Physical review}, vol.~36, no.~5, p. 823, 1930.

\bibitem{Luo2025}
H.~Luo \emph{et~al.}, ``{Digital Twin Aided Massive MIMO CSI Feedback: Exploring the Impact of Twinning Fidelity},'' \emph{IEEE Transactions on Communications}, 2025, early access.

\bibitem{Kazhdan2006}
M.~Kazhdan \emph{et~al.}, ``{Poisson Surface Reconstruction},'' in \emph{Proc. of Eurographics Symposium on Geometry Processing}, vol.~7, no.~4, 2006.

\bibitem{Remcom}
Remcom, ``{Wireless InSite},'' \url{http://www.remcom.com/wireless-insite}.

\bibitem{Alkhateeb2019}
A.~Alkhateeb, ``{DeepMIMO: A Generic Deep Learning Dataset for Millimeter Wave and Massive MIMO Applications},'' in \emph{Proc. of Inf. Theory and Appl. Workshop}, 2019, pp. 1--8.

\end{thebibliography}
